\documentclass{elsart}
\usepackage{graphicx}
\usepackage{bm}
\usepackage{hyperref}
\newif\ifcaption
\captiontrue

\def\qr{{\bf r}}

\def\qR{{\bf R}}

\def\qJ{{\bf J}}
\def\qbe{\begin{displaymath}}
\def\qee{\end{displaymath}}
\def\qbel#1{\begin{equation}\label{#1}}
\def\qeel{\end{equation}}
\def\qbea{\begin{eqnarray*}}
\def\qeea{\end{eqnarray*}}
\def\qbeal#1{\begin{eqnarray}\label{#1}}
\def\qeeal{\end{eqnarray}}

\begin{document}
\begin{frontmatter}

\title{Comparison of Rotational Energies and Rigidity of OCS-paraH$_2$ and
OCS-$^4$He complexes}

\author{R.~E.~Zillich and K.~B.~Whaley}

\address{Department of Chemistry and Pitzer Center for Theoretical
Chemistry, University of California, Berkeley, CA 94720}

\begin{abstract}
We analyze the nature of the rotational energy level structure of the OCS-He and 
OCS-H$_2$ complexes with a comparison of exact calculations to several different
dynamical approximations.
We compare with the  clamped coordinate quasiadiabatic approximation
that introduces an effective potential for each
asymmetric rotor level, with an effective rotation Hamiltonian
constructed from ground state averages of the inverse of the inertial
matrix, and investigate the usefulness of the Eckart condition to decouple
rotations and vibrations of these weakly bound complexes between linear
OCS and $^4$He or H$_2$.
Comparison with exact results allows an assessment of the accuracies of the different
approximate methods and indicates which approaches are suitable for larger clusters
of OCS with $^4$He and with H$_2$.
We find the OCS-H$_2$ complex is considerably more rigid than the OCS-He
complex, suggesting that semi-rigid models are useful for analysis of
larger clusters of H$_2$ with OCS.
\end{abstract}


\end{frontmatter}

\section{Introduction}

Until recently, Helium-4 has been the only known boson superfluid liquid.
Many aspects of superfluidity have been studied experimentally,
phenomenologically, and theoretically. These include its relation to Bose-Einstein
condensation, the temperature dependence of the superfluid
fraction in the 2-fluid model, the finite size-dependence and the influence
of atomic and molecular impurities
in $^4$He on the superfluid fraction. The latter
two issues have raised new, interesting and more basic questions. In particular,
what is the validity of
phenomenological models on a microscopic scale, and how should we detect
superfluidity of small systems. These questions ultimately demand a full
microscopic understanding of superfluidity.

Rovibrational IR spectroscopy experiments with chromophores in helium droplets
\cite{ToeViWha01,Toennies98review} have confirmed superfluid behavior for
systems as small as approximately 60 $^4$He atoms \cite{Toennies98} by
observing the free rotation of the OCS molecule.
Subsequently, a new field of research has been opened by similar experiments
in small para-hydrogen clusters embedded in low temperature helium
droplets.\cite{Toennies00}
Spectroscopic experiments on OCS in these hydrogen clusters show spectroscopic anomalies
which have been interpreted as indicating the existence of superfluidity of para-H$_2$ for
clusters small enough to be fluid-like rather than forming a solid
shell around the chromophore.  These conclusions have recently been verified
by path integral calculations showing that an anistropic superfluid fraction appears at low temperatures ($T < 0.3$~K).\cite{kwon02}
In addition, new experimental studies on very small complexes with less than
one solvation shell of $^4$He or H$_2$ obtained rotational/rovibrational
excitation energies for complexes of well-defined size, OCS-pH$_2$
and OCS-$^4$He$_N$,
$N=1,\dots,8$.\cite{higgins99,tang01,tang02,xujaeger01cpl,tang02science}
New theoretical work for small OCS-$^4$He$_N$ complexes \cite{paesani02} show an interesting transition from
semi-rigid rotation of the whole cluster for small $N$ to partial
decoupling of the $^4$He motion and the OCS rotation for increasing $N$.
As explained in detail in Ref.~\cite{paesani02}, this implies a transition
from a molecular complex to quantum solvation.

Quantitative microscopic calculation of the spectrum of excitation energies
of OCS-(pH$_2$)$_N$ and OCS-$^4$He$_N$
clusters is an important step towards the understanding of
the rovibrational spectra of these unusual clusters.
For $N\gg 1$, this is still an extremely demanding task with current methodological and
computational capabilities. Exact excitation energies are often not
accessible and it is therefore necessary to make use of dynamical
approximations, based whenever
possible on existing knowledge of {\em e.g.} small cluster structures or on microscopic
understanding of quantum solvation structures. 
In this paper, we therefore present
a survey of approximate methods for the calculation of rotational
energies of small Van der Waal complexes that have potential application to large
quantum clusters of OCS in $^4$He and in H$_2$. 
In order to be able to critically assess the
various approximations by comparing with exact results, we focus here on the smallest case, $N=1$.

The methods we employ here are all based on diffusion Monte Carlo (DMC) sampling. Our
focus on DMC methods derives from their suitability for application to larger sizes
that are beyond the scope of basis set approaches. 
DMC yields exact results for ground state energies and the approximations employed here
all take advantage of this. The approximate methods all assume
some degree of rigidity, an assumption which we expect would progressively
fail to be justified as one would go to larger clusters, with helium failing 
at smaller cluster size than hydrogen.  In section \ref{excstate}
we give an overview of how to calculate
approximate rotational excitations. We described the clamped coordinate quasiadiabatic
approximation of Quack and Suhm \cite{quackquack91}, a modification of it
which incorporates the Eckart condition \cite{eckart35} for semi-rigid
complexes, and a perturbative calculation of rotational
constants following Ref.~\cite{ernesti94}.
In section~\ref{results}, we compare these results with the exact bound
state energies determined by the program BOUND \cite{bound}. We discuss here also 
the relative rigidity of the OCS-He and OCS-H$_2$ systems as apparent from the
grouping of exact energy levels into (ro)vibrational and rotational subsets,
where ``vibrational'' from now refers to the radial motion of H$_2$ or
$^4$He with respect to OCS.
Section~\ref{sec:conclusion} summarizes and provides conclusions.

\section{Ground state}
\label{groundstate methodology}

The dimers OCS-pH$_2$ and OCS-$^4$He are weakly bound van der Waals
complexes which demand a full quantum mechanical calculation of the $^4$He degrees
of freedom, due to their
large zero-point motion. Exact methods \cite{collocation,leforestier94b,bound}
are available to determine
ground and excited states of the dimers. However, we are interested in larger clusters
OCS-(H$_2$)$_N$ and OCS-$^4$He$_N$, with special focus
on the influence of the superfluid properties of the solvation
shell around OCS \cite{grebenev00,kwon00}. For this reason, in this paper
we employ diffusion Monte Carlo (DMC) which is applicable to
large systems of sizes on the order of $N=100$, and compare with the result of exact 
calculations for $N=1$.

The theory and implementation of DMC has been described extensively
in the literature \cite{guardiolaValencia98,hammondbook,umrigar93}.
We employ here a mixed branching-weighting
algorithm. Reliable interaction potentials
$V$ between OCS and $^4$He, and H$_2$ (assumed spherical),
respectively, have been calculated by
Higgins {\it et.al.\/}~\cite{higgins99,higgins02}. Additional interaction potentials
are available for the OCS-He system~\cite{gianturco00,hutson01}.
Fig.~1 shows cuts of the OCS-H$_2$ and two OCS-He potentials along the
minimum energy path. In order to minimize computational cost in DMC,
all potentials were expanded in Legendre polynomials and the resulting
expansion coefficients interpolated with splines. This is responsible
for  the small irregularities in the lower panel of Fig.~1, but we have
verified that these do not affect our results.
In this work we will not be concerned
with the determination and quality of the potentials in regard to experimental
measurements, and shall therefore start our
investigation with the given Hamiltonian
\qbel{eq:H}
  H = -{\hbar^2\over 2m_X}\nabla_X^2
      -{\hbar^2\over 2M}\nabla_0^2
      -B\Big({\partial^2\over\partial\varphi_x^2}
           + {\partial^2\over\partial\varphi_y^2}
        \Big)
      + V(|\qr_X-\qr_0|,\cos\theta),
\qeel
where $m_X=m({\rm H}_2)$ and $m(^4{\rm He})$, $B$ is the rotational
constant of free OCS in its ground state, and $M$ its mass. $\partial\varphi_x$
and $\partial\varphi_y$ are the infinitesimal
angles of rotation of the principal axis frame of OCS, $r\equiv |\qr_X-\qr_0|$
is the distance between atom and center of mass of OCS,
and $\theta$ the angle between the OCS axis and the location of the
atom. 
We employ the rigid body diffusion Monte Carlo
algorithm\cite{buch92,viel02cpc.pdf} in which the rotational
degrees of freedom $\varphi_x$ and $\varphi_y$ are sampled by random walks in the
angular variable taken in the principal axis frame of the molecule.
Although it is not essential for a system of only 8 coordinates
$(\qr_X,\qr_0,\varphi_x,\varphi_y)$, of which only
$r$ and $u\equiv\cos\theta$ enter in the potential calculation, we employ
biased DMC here, sampling the state $\Phi_0\Phi_T$ instead of $\Phi_0$,
with the trial wave function
\begin{eqnarray}
  \Phi_T(r,u) = \exp&\Big[& a_1 r^\alpha + (t_1+\eta_1(u-u_1)^2)\log {r\over a_2}
		\nonumber \\
  &&
              +\ (t_2+\eta_2r^2(u-u_2)^2) e^{\eta-cr} \Big]
\label{eq:phit}
\end{eqnarray}
which is flexible enough also for $N>1$ complexes.
The parameters of $\Phi_T$ are chosen to approximately
minimize the energy expectation value, and are given in table~\ref{tab:par}.
This trial function has both angular and radial dependence, and thus it requires the
full rotational importance sampled algorithm developed in Ref.~\cite{viel02cpc.pdf}.
Importance-sampled DMC yields exact values for the ground state energy and also for
expectation values of operators commuting with the Hamiltonian. 
Unless otherwise specified, all expectation values of operators not
commuting with the Hamiltonian are computed
here with descendent weighting~\cite{liu74} implemented
according to the procedure for pure estimator of 
Ref.~\cite{boronat95a}.  This corrects the mixed expectation values
that result from importance sampling ({\em i.e.}, averages over the
product $\Phi_0\Phi_T$) by a factor of $\Phi_0/\Phi_T$ to obtain the exact 
expectation value over $\Phi_0^2$.  This is particularly important for
structural quantities.

\section{Methods for Rotational excitations}
\label{excstate}

Since DMC transforms the Schr\"odinger equation to a diffusion equation
whose asymptotic solution is the ground state, DMC is often very efficient for
the calculation of ground state properties while imposing only a small
well controllable time step bias. The situation is quite different for excited
states. Exact calculation of excitation energies
is possible with the projection operation imaginary time spectral evolution approach
(POITSE) which is generally very time-consuming.\cite{blume97,viel01} Alternatively,
excited states can be calculated with fixed node approximations, but these generally
introduce an unknown amount of bias by the choice of nodal surfaces\cite{viel01}. The 
methods described below share the element of approximation with fixed node
in the sense that they are approximative, but possess the advantage that
it is very straightforward and can be implemented with the same efficiency
as ground state energy calculations.

\subsection{Clamped coordinate quasiadiabatic approximation (ccQA)}
\label{ssec:ccQA}

We summarize here the procedure which Quack and Suhm \cite{quackquack91} 
introduced to calculate rotational levels of the van der Waals dimer (HF)$_2$.
The rigid-rotor Hamiltonian $H_{\rm rot}$ of the instantaneous configuration
$\qR$ at diffusion time $t$
is obtained by diagonalizing the associated instantaneous inertial tensor
$I$. For OCS-X$_N$, its elements in the body fixed frame of OCS are
$$
  I_{\alpha\beta} =
    {\hbar^2\over 2B} (\delta_{\alpha,\beta}-\delta_{\alpha,3}\delta_{\beta,3})
  + \sum_i m_i (\delta_{\alpha,\beta} r_i - x_i^\alpha x_i^\beta )
$$
with $r_i=\sum_\gamma x_i^\gamma x_i^\gamma$.
Diagonalization yields principal values $I_A$, $I_B$, and $I_C$, and hence
the rotational constants $A=\hbar/8\pi^2cI_A$, $B=\hbar/8\pi^2cI_B$, $C=\hbar/8\pi^2cI_C$.
For $N=1$ these should correspond to those of an asymmetric top for both
He and H$_2$ complexes, as seen experimentally~\cite{tang01,tang02}. 
For given total angular momentum $J$, the resulting rigid-rotor cluster Hamiltonian
is
\qbel{eq:ccQA1}
  H_{\rm rot} = A\hat J_x^2 + B\hat J_y^2 + C\hat J_z^2.
\qeel
This is diagonalized in a symmetric rotor basis,\cite{zare}
resulting in eigenenergies $\epsilon_{J,\tau}$, $\tau=-J,\dots,J$.
Since $H_{\rm rot}$
depends parametrically on the configuration
$\qR$, the energies $\epsilon_{J,\tau}$ are also
parametrically dependent on $\qR$. As pointed out by Quack and Suhm
in Ref.~\cite{quackquack91}, they can therefore 
be regarded as providing effective centrifugal potentials,
$V_{J,\tau}(\qR) = \epsilon_{J,\tau}$, which may be added to the cluster
Hamiltonian (\ref{eq:H}) to yield an effective Hamiltonian for different 
rotational states of the cluster as a whole:
\begin{equation}
H_{J,\tau}(\qR) = H(\qR) + V_{J,\tau}(\qR).
\label{eq:H_Jtau}
\end{equation}
Because the Hamiltonians $H(\qR)$ and $H_{J,\tau}(\qR)$ differ only by
the centrifugal potential terms, implementation of correlated
sampling \cite{wells85,schinacherdiss} on all systems $H_{J,\tau}(\qR)$
with a single random walk, but employing different individual weights
is comparatively straightforward. 
Each system results in an excitation energy $E_{J,\tau}$, where these are
the ground state energies of the respective effective Hamiltonians.
The ground
state energy $E_0$ can be calculated inexpensively together with 
all excitation energies $E_{J,\tau}$ in the single DMC run.

\subsection{Eckart frame ccQA (EccQA)}

The diagonalization of the moment of inertia tensor $I$ at every (imaginary time) 
instant $t$ amounts to choosing
the principal axis of $I$ as the instantaneous molecule fixed axis
system, which thus moves with the motion of the generalized coordinate $\qR$. However, 
the axis system which
separates rotational and small vibrational motion most effectively
is the Eckart frame \cite{eckart35} and this is the preferred
axis system for the derivation of rovibrational Hamiltonians.
Therefore, we modify the above ccQA prescription in the following manner to 
accommodate the preferred Eckart reference frame.
\begin{enumerate}
\item
 Instead of diagonalizing the instantaneous
 inertial tensor $I$ (which is of course independent of the axis system
 used), we calculate the components $I_{i,j}$ in the Eckart frame and
 invert this matrix.
\item
 Instead of solving for the eigenvalues of the asymmetric top
 Hamiltonian (\ref{eq:ccQA1}),
 we solve for the eigenvalues $\epsilon_{J,\tau}$ of
 \qbe
   H_{\rm rot} = 
   \sum_{i,j}(I^{-1})_{i,j} \hat J_i \hat J_j
 \qee
 This Hamiltonian is also used in the method described in the next section.
\end{enumerate}

For the determination of the instantaneous Eckart frame, a reference
geometry needs to be specified. In the case of OCS-X, it can be
specified by the position of X with respect to the OCS center of
mass and molecular axis, $r_{\rm ref}$ and $\cos\theta_{\rm ref}$.
As reference values of these coordinates, we choose the ground state expectation values
$r_{\rm ref}=\langle r\rangle$ and
$\cos\theta_{\rm ref}=\langle\cos\theta\rangle$. For OCS-H$_2$
and OCS-$^4$He we find these expectation values are given by
$r_{\rm ref}=3.704{\rm \AA}$ and $3.935{\rm \AA}$, and
$\theta_{\rm ref}=105.7^\circ$ and $108.4^\circ$, respectively.

\subsection{Rotational constants from ground state averages (GSA)}
\label{gsa}

For semi-rigid molecules, rotational and vibrational degrees of
freedom can be treated in a Born-Oppenheimer approximation
in which the Hamiltonian is averaged over the rotational ground state.
This is achieved by defining
an effective rotational Hamiltonian $H_{\rm rot}^{\rm eff}$
\qbel{eq:Heff}
  H_{\rm rot}^{\rm eff} = 
  \sum_{i,j}\langle\Psi_0 | (I'^{-1})_{i,j}|\Psi_0\rangle \hat J_i \hat J_j
\qeel
where $I'_{i,j}$ are components of the effective inertial
tensor (see Refs. \cite{ernesti94,WilsonDeciusCross})
calculated in the Eckart frame $(x,y,z)$. We use the definition of
of the Eckart frame axes given in Ref.~\cite{ernesti94}, with interchange
of $y$ and $z$.
Since for small amplitude vibrations $I'$ differs only slightly from the
pure inertial tensor $I$, we have followed
Ref.~\cite{ernesti94} and also calculated
$\langle\Psi_0 | (I^{-1})_{i,j}|\Psi_0\rangle$. For OCS-X, the
effective rotational Hamiltonian (\ref{eq:Heff}) takes the form
\qbel{eq:Heff1}
  H_{\rm rot}^{\rm eff} = 
    A \hat J_x^2 + B \hat J_y^2 + C \hat J_z^2
  + D(\hat J_x \hat J_y + \hat J_y \hat J_x)
\qeel
where $y$ is defined as the axis perpendicular to the OCS-X plane. The explicit
expressions for
$A$, $B$, $C$, and $D$ are given in Ref. \cite{ernesti94}.
Rotational energies are then obtained by expanding $H_{\rm rot}^{\rm eff}$ in
the symmetric rotor basis and diagonalizing the matrix. 
As noted already above in Section~\ref{groundstate methodology},
we compute the exact ground state expectation values in Eq.~(\ref{eq:Heff}) 
by descendant weighting of the importance sampling, in order to
remove the trial wave function bias.

By an {\it ad hoc\/} combination of ccQA and GSA,
one can also calculate rotational constants
from (approximate) excited state averages in a similar fashion and use these
to go back and
calculate rotational energies. The procedure is as follows. First,
sample an approximate excited
state corresponding to $(J,\tau)$ defined by the effective Hamiltonian 
Eq.~(\ref{eq:H_Jtau}) of the ccQA approach
explained in section~\ref{ssec:ccQA}. Second,
calculate the rotational constants $A$, $B$, $C$, $D$ as averages over these
approximate excited states, {\em i.e.}, 
$\langle\Psi_{J,\tau} | \hat A|\Psi_{J,\tau}\rangle$ etc.  This needs to be done also
with descendent weighting to correct for the trial function bias. Third,
diagonalize $H_{\rm rot}^{\rm eff}$ to obtain an updated estimate
of the excitation energy $E_{J,\tau}$. We found that this {\it ad hoc\/} estimate
turned out to be inferior to the ccQA results for both complexes, OCS-(H$_2$) and
OCS-$^4$He.

\section{Results}
\label{results}

We have applied the DMC-based approximate methods described above to
the calculation of rotational excitation energies and rotational constants
for OCS-(H$_2$) and OCS-$^4$He and compared with the exact energies
which we have determined using the program BOUND \cite{bound}.
To relate the exact reference energies to rotational constants, we fitted the BOUND 
rotational energies to the eigenenergies
of the Watson A-reduced Hamiltonian.\cite{BunkerJensen} 
In case of the GSA calculation, the resulting rotational parameters
$A,B,C,D$ are used to construct the effective Hamiltonian~(\ref{eq:Heff1}) which we
then diagonalized to obtain rotational energies. 

In all calculations reported here, the imaginary time step value in the DMC
walk was $dt/\hbar=0.00011{\rm cm}$ for
OCS-(H$_2$) and $dt/\hbar=0.00022{\rm cm}$ for OCS-$^4$He.

Fig.~2 shows the respective ground state
densities of H$_2$ and $^4$He with respect to OCS,
$\rho(r,\theta)$. These densities are obtained by simply
binning the coordinates of H$_2$ and $^4$He in the body fixed frame of OCS
in an unbiased DMC calculation, {\it i.e.} with
trial function $\Phi_T=1$, and then squaring the amplitude. 
The densities are qualitatively quite similar for the two complexes, with
the helium density being somewhat more delocalized. 
In particular, we note that $^4$He has a non-negligible
probability to reside near the south pole of OCS (the oxygen end, at negative $z$). 
This leads us to expect that a semi-rigid approximation for 
OCS-$^4$He might be problematic.

\subsection{Exact excitation energies from BOUND}
\label{results_bound}

Fig.~3 summarizes the exact and quasi-adiabatic energy levels for OCS 
complexed with H$_2$ (top panel), and for OCS complexed with $^4$He (middle and bottom
panels).  The exact energy levels are shown as horizontal bars.
For the helium complex, we show the results from two different OCS-He 
potentials, namely the MP4 potential of Higgins and Klemperer~\cite{higgins99} (middle
panel) and the HHDSD potential of Gianturco and Paesani~\cite{gianturco00}
(bottom panel).
The ground state energy of OCS-H$_2$ is -74.59 cm$^{-1}$ and of OCS-$^4$He
is -16.38 cm$^{-1}$ (MP4) / -15.85 cm$^{-1}$ (HHDSD).
The exact results for OCS-H$_2$ show a very well-defined 
energetic separation between the low energy rotational states and the higher energy
states associated with the vibrational motion of the H$_2$ molecule in the OCS-H$_2$
potential.  For low values of $J$, the total angular momentum, these two different sets of
states are separated by $\sim$ 25 cm$^{-1}$. This separation decreases as $J$ increases
and the rovibrational and purely rotational states are expected to mix, but even for $J=6$, the highest
value shown here, there is still a clear separation of $\sim$ 5 cm$^{-1}$ between the highest rotational sublevel and the first rovibrational level.

These results indicate that OCS-H$_2$ can be described as a rigid rotor and suggests that
dynamic approximations based on such an assumption will provide a good starting point
for {\em e.g.}, analyzing energy levels in larger clusters.
This situation changes considerably for the lighter OCS-He complex.  
With both the MP4 potential (middle) and even more so with the HHDSD potential (bottom) 
we find the energy separation between rotational and vibration levels greatly 
reduced.  For small $J$, this separation is only $\sim$ 8 cm$^{-1}$ with the MP4 potential,
and further decreases to $\sim$ 4 cm$^{-1}$ with the HHDSD potential.
This smaller separation between the purely rotational and rovibrational levels
for the helium complex can be ascribed to the weaker interaction
between OCS and $^4$He.

The energetic spread of the rotational sublevels $\tau$
is also different between the two systems,
with the helium complex levels being compressed relative to the corresponding
levels in the hydrogen complex levels: the lighter mass of
H$_2$ results in a large rotational constant $A$ compared to $B$ and $C$,
which leads to a steeper increase of the rotational energy with
increasing $\tau=K_a-K_c$, {\em i.e.\/} for large $K_a$, which approximately
corresponds to the fast rotation of H$_2$ about a tumbling OCS.

However, this smaller spread for the helium complex is clearly outweighed
by the decrease in rovibrational energies.
As a result of the smaller gap between rotational and rovibrational
energies for OCS-He, a mixing of rotational and rovibrational levels is predicted
at $J\geq4$ for the HHDSD potential (bottom panel), and at $J \geq 5$ for the
MP4 potential (middle panel).  In comparison, for the OCS-H$_2$ complex
the mixing does not occur until $J=7$.  The onset of this mixing of
rovibrational and rotational levels at small $J$ values is another indicator of non-rigid, ``floppy'',
behavior, seen here in the extreme for the helium complex. Therefore,
from energy consideration,
pure rotations cannot be distinguished from rovibrations
at large values of the total angular quantum number $J$.

Table~\ref{tab:1} and \ref{tab:2} list the parameters of the
Watson A-reduced Hamiltonian obtained by fitting the energy levels obtained
from the BOUND calculations to the following Watson A-reduced form: 
\begin{eqnarray}
  H_A &=& {1\over 2}(B+C)\qJ^2 + [A-{1\over 2}(B+C)]\qJ_a^2
      + {1\over 2}(B-C)(\qJ_b^2-\qJ_c^2) - \Delta_J\qJ^4
  \nonumber\\
    &&- \Delta_{JK}\qJ_a^2\qJ^2 
    - \Delta_K\qJ_a^4 - 2\delta_J\qJ^2(\qJ_b^2-\qJ_c^2)
  \nonumber\\
    &&- \delta_K[\qJ_a^2(\qJ_b^2-\qJ_c^2)+(\qJ_b^2-\qJ_c^2)\qJ_a^2].
  \label{eq:Watson}
\end{eqnarray}
For comparison we show the corresponding experimental values of these
parameters as determined spectroscopically in Refs.~\cite{tang01} and
\cite{tang02}. These were obtained by fitting to energy levels
reaching up to $J=6$, but still lying below the rovibrational levels.
These tables provide a reference for the rotational constants derived from
the approximate methods, described below, and provide a measure of the
accuracy of the three potentials~\cite{higgins99,higgins02,gianturco00}
considered in this work. Both complexes are fit by asymmetric top parameters.

\subsection{Rotational energies from ccQA and Eckart modification (EccQA)}
\label{results_ccQA}

Both the assumption of separation of vibrational and rotational motion
that is central to the ccQA approach, and the resulting effective Hamiltonian
(\ref{eq:Heff}) break down when the respective vibrational and rotational energies
are of similar magnitude.  Thus from Fig.~3 and the discussion of the
exact energies above, it appears that ccQA approach should break down 
at $J \sim 4-5$ for the helium complex, and at $J \sim 7$ for the hydrogen complex. 
The GSA calculation is also expected to fail at these $J$ values.
Furthermore, neither of these approaches can
yield the higher rovibrational states, where, in addition to the rotation
of the complex, the H$_2$ or $^4$He atom also vibrates radially with respect
to the OCS molecule. To obtain estimates for the energies of
these excited states possessing both vibrational and rotational character, one would need 
to sample the various excited vibrational states for
$J=0$ ({\em e.g.}, by imposing nodal constraints on the trial wave 
function~\cite{lewerenz96,quackquack91},
and then add the centrifugal potential $V_{J,\tau}(\qR)$
to the Hamiltonian (\ref{eq:H}) as described above to obtain
vibrationally excited states with
$J>0$. Since this would introduce
more approximations, we refrained here from trying to obtain the
rovibrational excitation energies with ccQA
beyond the pure rotational energies.

The ccQA energies for OCS-H$_2$ and OCS-He are shown as crosses in 
Fig.~3.  They generally track the corresponding asymmetric top
rotational energy levels, showing the greatest accuracy for the more rigid
OCS-H$_2$ complex, and larger deviations for the more floppy OCS-He complex.
For comparison, we have plotted as a continuous dotted
line the quadratic behavior of the energy levels of free OCS, {\em i.e.}, 
$BJ(J+1)$, with $B=0.2028$ cm$^{-1}$.~\cite{grebenev00}
It is notable that while both complexes are asymmetric tops,
the free molecule line does nevertheless track the lowest set of
rotational levels for OCS-H$_2$, while for OCS-He the lowest rotational levels lie
considerably lower than the free molecule values.
In Figs.~4a, 4b, and 5,
we show the error of the ccQA and EccQA calculations
of rotational levels with respect to the exact BOUND energies,
$\Delta E_{J,\tau}=E_{J,\tau}({\rm (E)ccQA})-E_{J,\tau}({\rm BOUND})$,
for values of $J$ where pure rotational levels can still be identified.
Not surprisingly, the errors introduced by the (E)ccQA approximation
are larger for the less rigid OCS-$^4$He complex,  see the larger $y$
scale in Fig.~5, relative to Fig.~4a.
Nevertheless, the deviations  of (E)ccQA from the exact
energies are still quite small in both cases, considering that both dimers, especially
OCS-$^4$He, experience large zero-point motion (see Fig.~2).
Comparing now the original quasi-adiabatic approach (ccQA, closed symbols
in Figs.~4a, 4b, and 5) with the Eckart
modification (EccQA, open symbols), we see that the introduction of the
Eckart reference frame does not improve the accuracy of the results.
For completeness, we also have compared the relative errors 
for OCS-He when the two different potentials are used, MP4 and HHDSD (not
shown).  Since the HHDSD interaction is less anisotropic, Fig.~1,
and allows for larger
zero-point motion of the $^4$He atom, the (E)ccQA approximation
yields poorer results with th HHDSD interaction than with
the MP4 interaction, by up to a factor of two for $J=1-3$.

Fig.~6 shows the mean distance $\langle r\rangle$ of
H$_2$ and $^4$He, respectively, from OCS for the rotational excited
states $(J,\tau)$, obtained in the ccQA approximation.
Although $\langle r\rangle$ spans a smaller range for OCS-H$_2$ 
due to the larger rigidity, $\langle r\rangle$ rises more steeply
with $\tau=K_a-K_c$, because of the high energies involved with
the quantum number $K_a$ (see the comment above about the larger spread in
energies evident in Fig.~3).

\subsection{GSA}

The four rotational parameters of the ground state effective Hamiltonian
$H_{\rm rot}^{\rm eff}$, eq.~(\ref{eq:Heff1}), are tabulated
in Table~\ref{tab:four}. Comparison with the fits 
to the exact energy levels shown in Tables~\ref{tab:1} and \ref{tab:2} 
show that using the pure inertial
tensor $I$ instead of the effective inertial tensor $I'$ tends to improve
the agreement with the values of the asymmetric top
rotational constants $A,B,C$ from fitting to the
BOUND results. We note that, other than for the ccQA approximation,
GSA rotational constants are sensitive to the choice of axis system,
and the Eckart conditions may not be neglected.

The deviations of the eigenenergies of $H_{\rm rot}^{\rm eff}$ from the exact
rotational energies are shown in
Figs.~7a, 7b, and 8 for OCS-(H$_2$) and
OCS-$^4$He, respectively. Comparison with Figs.~4a and 4b
shows that, for OCS-H$_2$, the GSA deviations are of similar
magnitude as those of the ccQA approximation, with GSA doing
slightly better for smaller $J$ than ccQA and vice versa for larger $J$.
For OCS-$^4$He, the deviations of GSA increase faster with $J$.
Also, table~\ref{tab:four} shows that, just like ccQA, GSA is
less accurate for OCS-$^4$He than for OCS-(H$_2$). There is a
small, but finite probability for the $^4$He atom to be close to the
oxygen side, along the OCS axis, {\it i.e.\/} OCS-$^4$He
vibrating end-over-end. For these configurations, the rotational constants
$A$, $C (C')$, and $D$ diverge, leading to poor statistics in
the DMC calculation.

\section{Conclusions}
\label{sec:conclusion}

We have compared the exact rotational excitation energies of
the weakly bound Van der Waals complexes OCS-paraH$_2$ and OCS-$^4$He
with two approximate methods, namely, the clamped coordinate
quasiadiabatic approximation (ccQA) and ground state averages of the rotational
constants (GSA), as well as with variants of each of these methods. The purpose of this
survey is to assess which approximate methods will work best for larger
complexes OCS-(H$_2$)$_N$ and OCS-$^4$He$_N$, in the absence of non-stochastic 
exact techniques.

While for the GSA approach, fulfilling the Eckart condition
improves the results considerably, we found no significant improvement of
the ccQA results when the instantaneous inertial frame is replaced by the
Eckart frame (EccQA).
This is good news for larger size clusters, since
incorporating the Eckart condition brings with it the need to define a reference
structure, which would become increasingly cumbersome and less tractable for
growing $N$.
We also found that incorporation of a centrifugal correction
in the instantaneous inertial matrix in the GSA method (Section~\ref{gsa})
actually
tends to degrade the accuracy of the excitation energies. This
is in agreement with the findings of Ref.~\cite{ernesti94}.

A significant physical feature that emerged from comparison of the complete
energy spectrum of rotational and rovibrational states for the two complexes of
OCS with H$_2$ and with He, is the greater rigidity of the hydrogen complex, as
evident from the larger separation between purely
rotational and rovibrational states.
This was evident from surveying the overall pattern of energy levels,
Fig.~3. Experimental investigation of vibrationally
excited states for OCS-H$_2$ and OCS-$^4$He
would be useful, because their energies are most sensitive
to the potential surface $V(r,\cos\theta)$ at higher energies and larger
$r$ values.
This differentiation in rigidity apparent from examination of the excited
state spectrum is particularly interesting given the close similarity of the 
ground state densities for the two complexes (Fig.~2). 

A detailed projection for the breakdown of the assumption of separability of
vibrational and rotational excitations in the
multi-particle complexes OCS-X$_N$, $N>1$, is of course difficult, because
exact results for {\em all} excited state energies are currently impossible to
obtain for large $N$. However, the projection operator imaginary time
spectral evolution approach (POITSE) can provide exact energies
given a good trial wave
function. It yields the energies of {\it selected} rotational states, where
the selection is determined
by the choice of a projector acting on the ground state. POITSE calculations \cite{paesani02} of
rotational excitations of OCS-$^4$He$_N$ for $N\le 20$ show approximate
agreement with ccQA results for $N\le 5$ and $J=1$, with the ccQA results lying
always somewhat lower than the POITSE results. This agreement with
exact calculations indicates that the rigid coupling approximation underlying the
ccQA approximation is good in this regime.  For larger $N$, 
the POITSE rotational energy saturates in accordance with
experimental results, while the energies obtained in ccQA and GSA
monotonically decrease for increasing $N$.
This implies that 
not all $^4$He atoms in the first solvation shell can rigidly follow the OCS rotation,
as discussed extensively in Ref.~\cite{kwon00} and demonstrated explicitly
in Ref.~\cite{paesani02}.

\section{Acknowledgments}

This work was supported by the NSF under grant CHE-0107541,
and the Miller Institute for Basic Research in Science at the University
of California, Berkeley.
The authors would like to thank Francesco Paesani for providing
the program to fit the rotational energies to the Watson A-reduced Hamiltonian,
Patrick Huang for valuable discussions about correlated Monte Carlo
sampling, and E. Krotscheck for providing computational resources of
the Central Information Services of the Kepler University, Linz, Austria.

\ifcaption
\else
\pagestyle{empty}
\fi

\newpage
\bibliography{ocshehy,Mainbirgitt}
\bibliographystyle{elsart-num}
\newpage
\begin{figure}[ht]
\centerline{
  \includegraphics[width=0.8\linewidth]{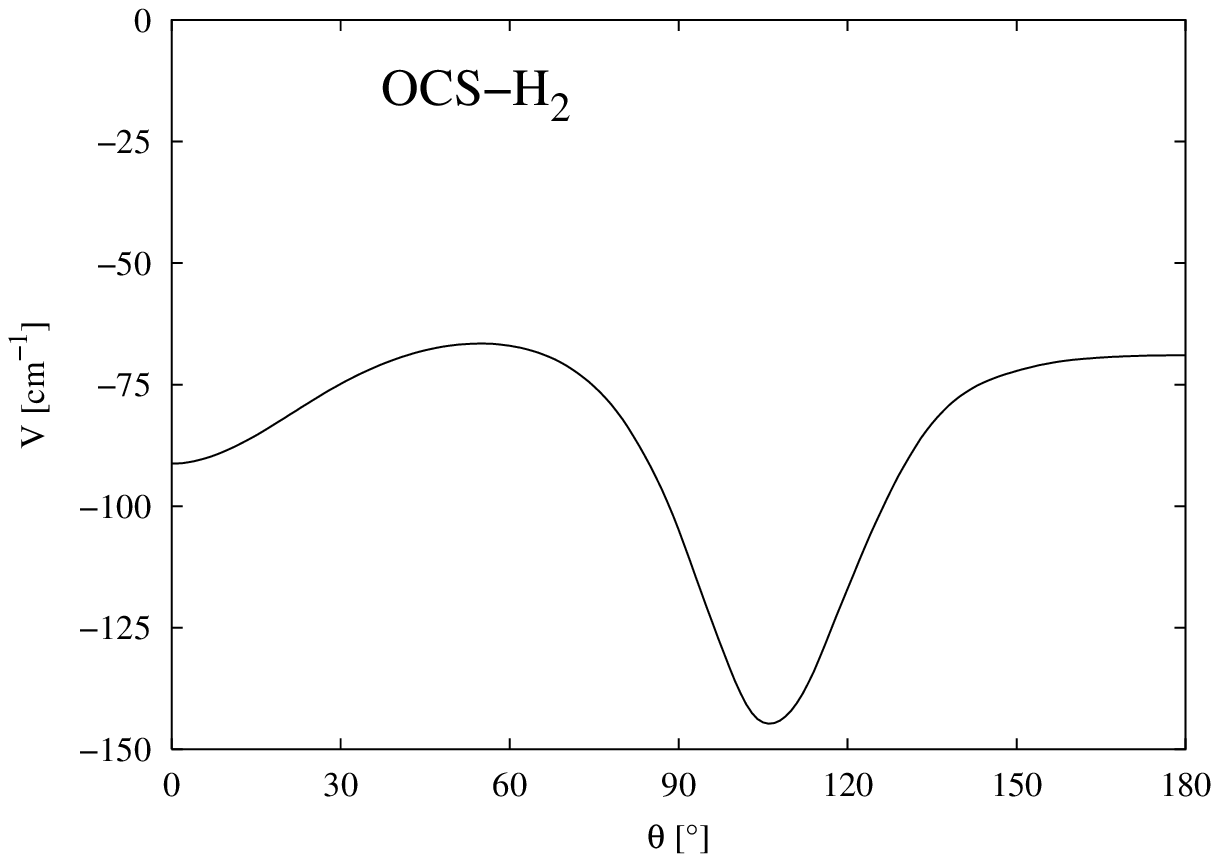}
}
\centerline{
  \includegraphics[width=0.8\linewidth]{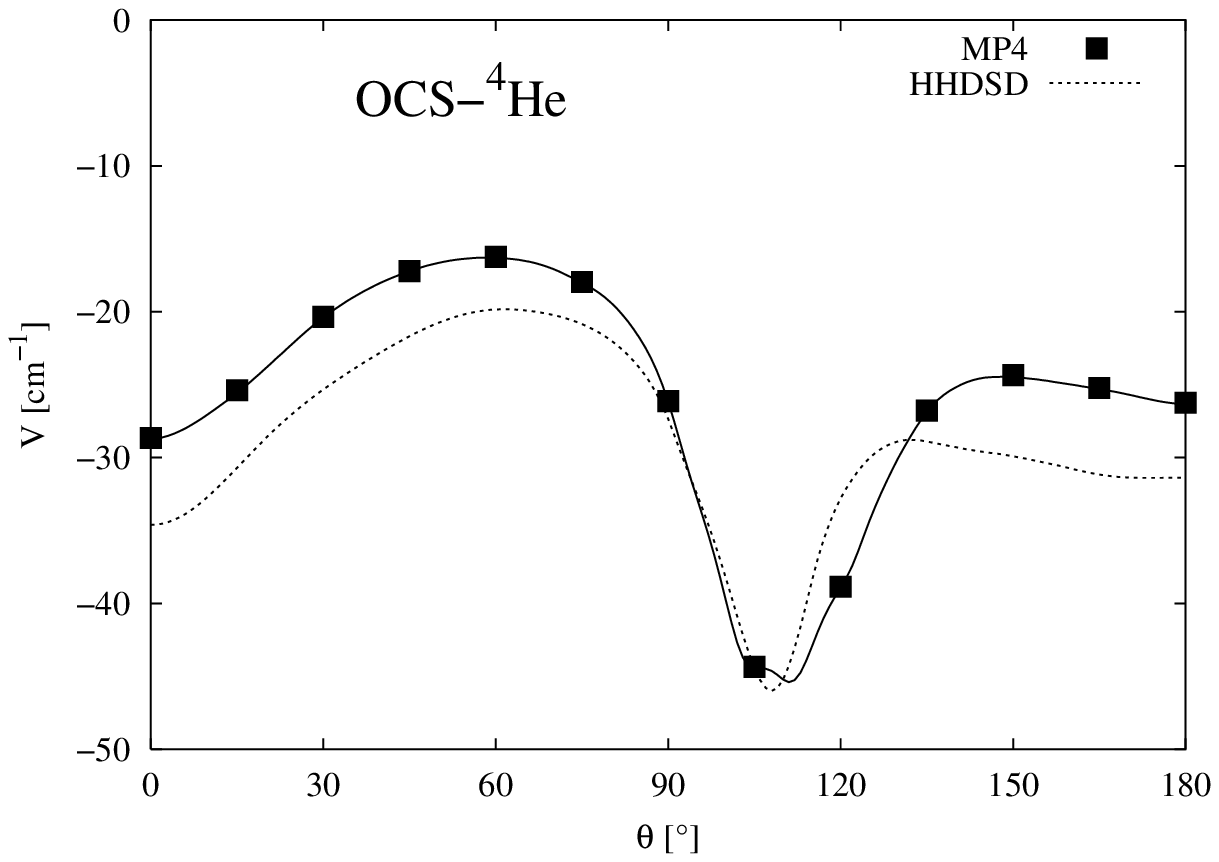}
}
\end{figure}
\ifcaption
        \noindent Fig.~1. 
        The interaction energy along the minimum energy path of the
        potential energy surface $V(r,\cos\theta)$ between H$_2$ and
        OCS (top)~\cite{higgins02} and $^4$He and OCS (bottom).
        For the latter we show both potentials used in our
        calculations, MP4 by Higgins {\it et.al.\/}~\cite{higgins99} (squares are
        ab initio points from that reference) and
        HHDSD by Gianturco {\it et.al.\/}~\cite{gianturco00} (dashed line).
        For usage in
        the DMC simulations, all potentials have been interpolated from
        the {\em ab initio\/} data points by
        splines in radial direction and by expansion in Legendre
        polynomials in angular direction, see Ref.~\cite{gianturco00}.
\fi
\clearpage
\newpage
\begin{table}
\caption{
        Parameters for the trial wave function (\ref{eq:phit}) where $r$ is
	in \AA.
\label{tab:par}
}
\bigskip
\begin{center}
\renewcommand{\arraystretch}{1.25}
\begin{tabular}{c|rrrrrrrrrrr}
\hline
& $a_1$ & $\alpha$ & $t_1$ & $\eta_1$ & $u_1$ & $a_2$ & $t_2$ & $\eta_2$ & $u_2$ & $\eta$ & $c$ \\
\hline
OCS-(H$_2$) & -6.6 & 0.79 & 6.4  & -1.28 & -0.22 & 0.529 & -1.0 & -0.71 & -0.22 & 5.9 & 1.59 \\
OCS-$^4$He  & -5.9 & 0.79 & 7.68 & -0.32 & 0.0   & 0.529 & -1.0 & -0.57 & -0.26 & 6.3 & 1.78 \\
\hline
\end{tabular}
\end{center}
\end{table}
\clearpage
\newpage
\begin{table}
\caption{
        Equilibrium configuration and root mean square deviations.
\label{tab:eq}
}
\bigskip
\begin{center}
\renewcommand{\arraystretch}{1.25}
\begin{tabular}{c|cccc}
\hline
& $\langle r\rangle$[\AA] & $\langle\cos\theta\rangle$ & $\Delta r$[\AA] & $\Delta\cos\theta$ \\
\hline
OCS-(H$_2$) & 3.704 & -0.270 & 0.364 & 0.158 \\
OCS-$^4$He  & 3.935 & -0.315 & 0.478 & 0.200 \\
\hline
\end{tabular}
\end{center}
\end{table}
\clearpage
\newpage
\begin{figure}[ht]
\centerline{
  \includegraphics[width=0.9\linewidth]{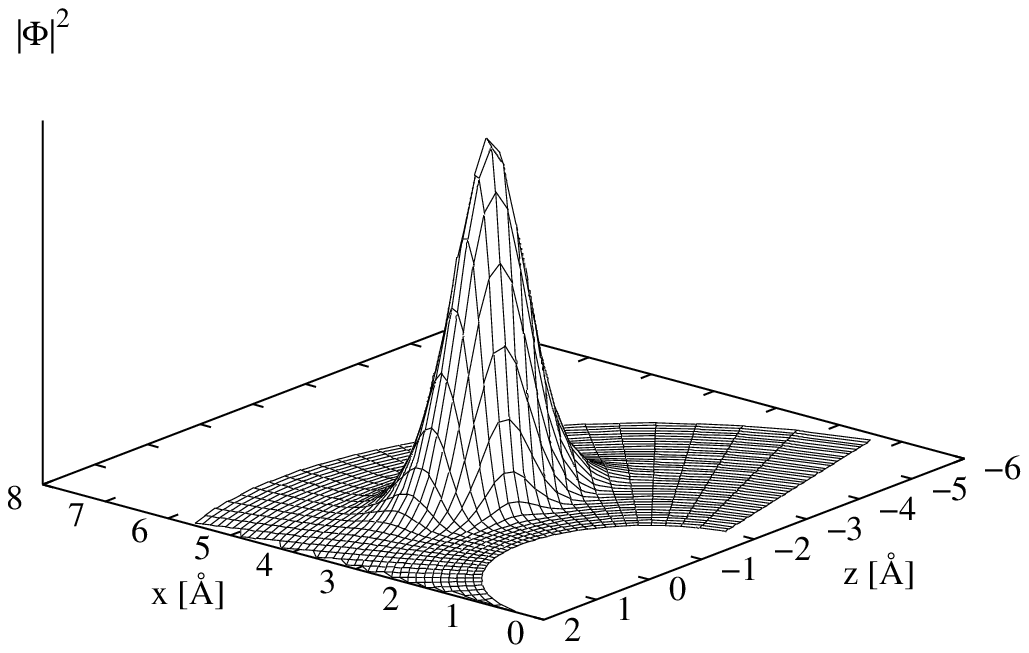}
}
\centerline{
  \includegraphics[width=0.9\linewidth]{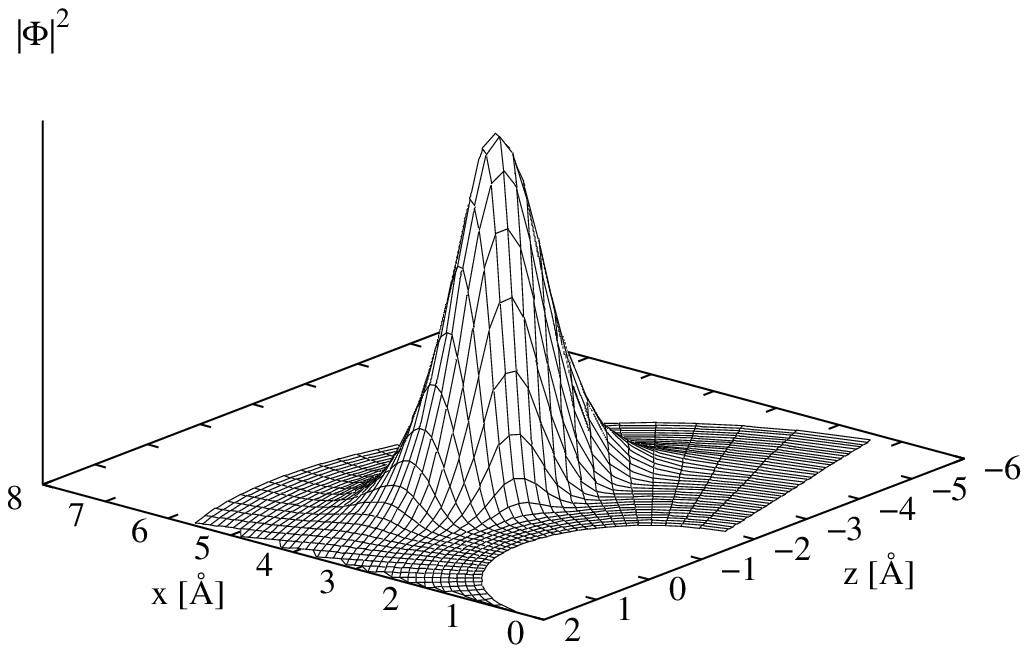}
}
\vspace {1truecm}

\ifcaption
  \noindent Fig.~2. 
  Probability density $\rho(r,\theta)$ for H$_2$ (top) and $^4$He
        (MP4 potential of Higgins {\it et.al.\/}~\cite{higgins99}, bottom) 
  to be found at distance $r$ from the OCS center of mass and
  at angle $\theta$ in the OCS fixed coordinate system, obtained from
  unbiased DMC.
  The OCS molecule is oriented along the $z$-axis, with oxygen at negative $z$, and
the $x$-axis is defined as any axis perpendicular to $z$. 
\fi

\end{figure}
\clearpage
\newpage
\begin{figure}[ht]
\centerline{
  \includegraphics[width=0.7\linewidth]{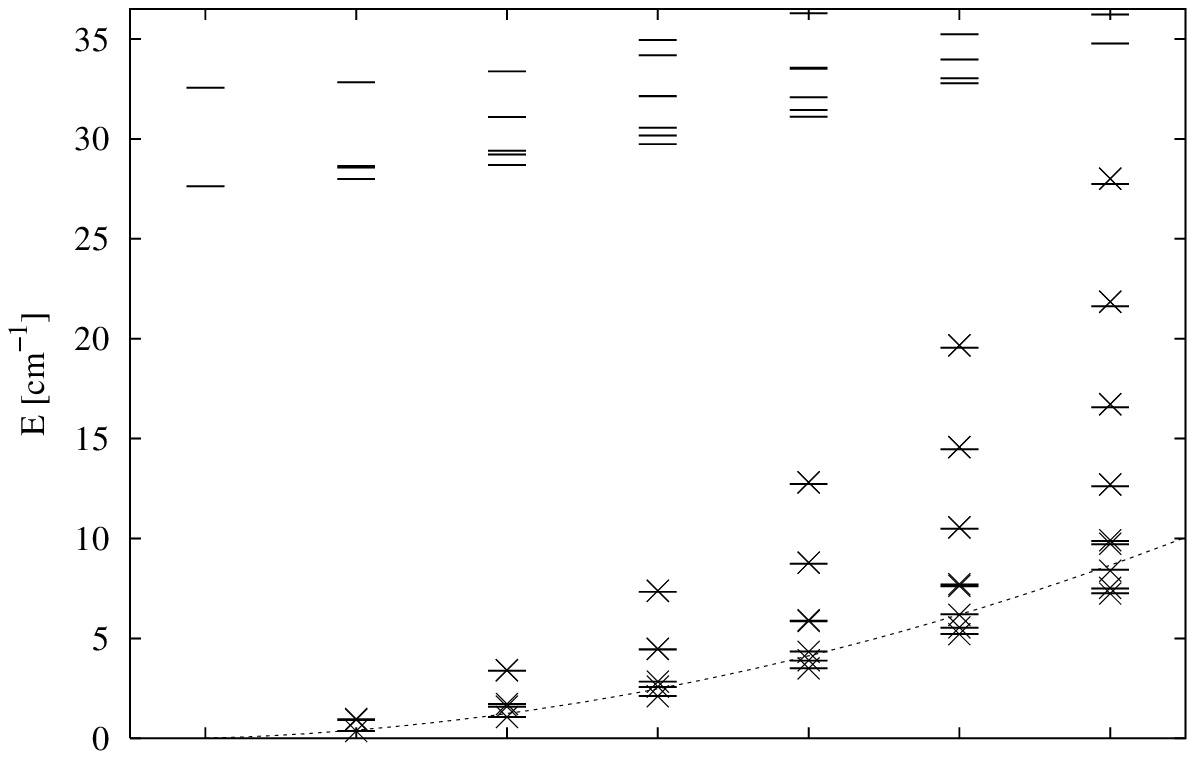}
}
\centerline{
  \includegraphics[width=0.7\linewidth]{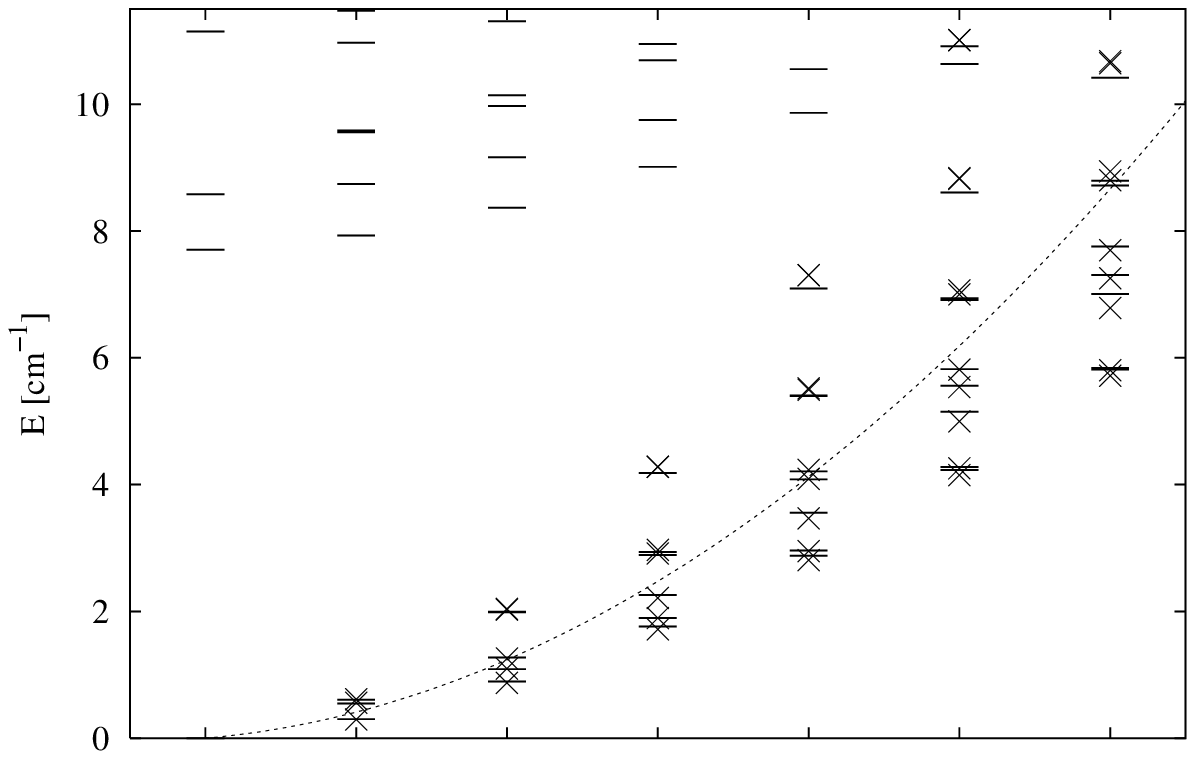}
}
\centerline{
  \includegraphics[width=0.7\linewidth]{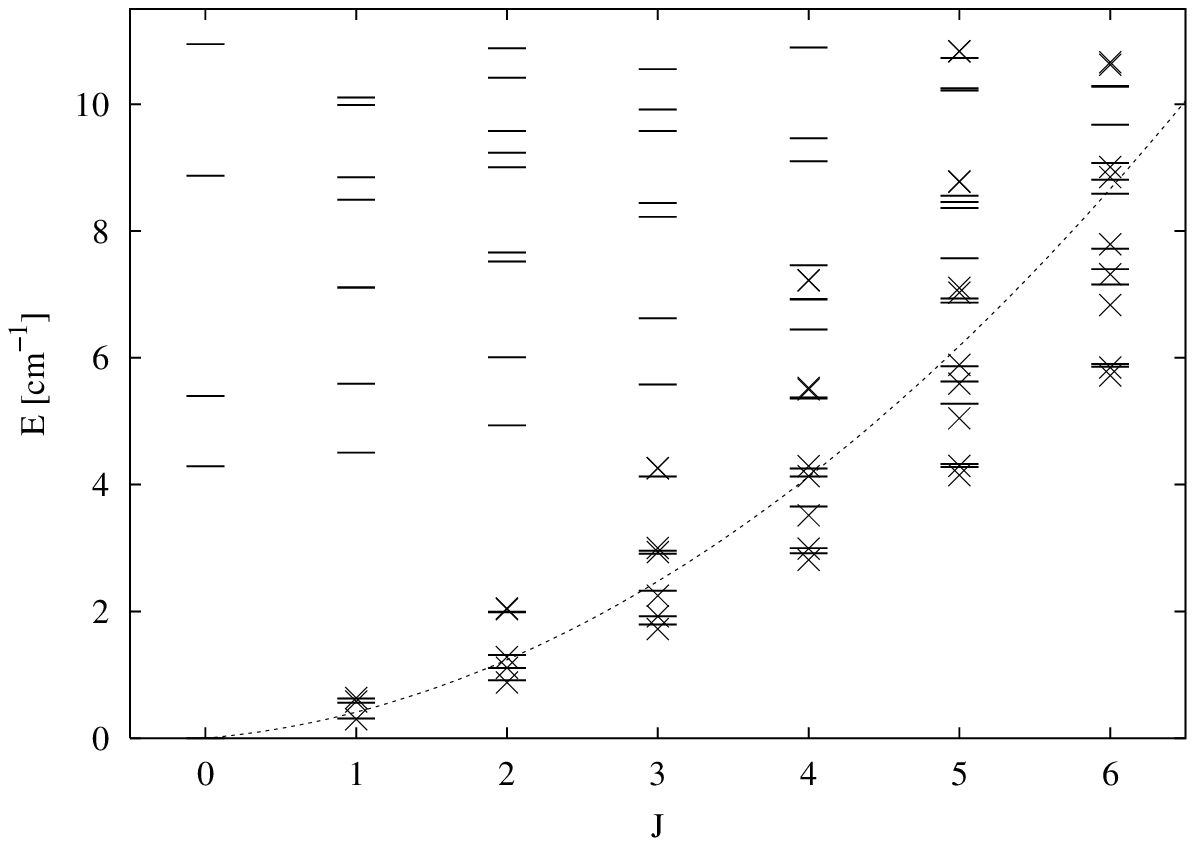}
}
\end{figure}
\vspace {0truecm}

\ifcaption
        \noindent Fig.~3. 
	Top panel:
        excitation energies $E_{J,i}$ of OCS-(H$_2$), obtained by
        BOUND and in the clamped coordinate
        quasi-adiabatic approximation; middle and bottom panels: $E_{J,i}$ for
        OCS-$^4$He, using the MP4 potential of Higgins {\it et.al.\/}~\cite{higgins99}
        (middle) and the HHDSD potential of \cite{gianturco00} (bottom).
        For comparison, the dotted line shows the quadratic behavior of the
        free OCS rotational energy levels $BJ(J+1)$, with
        $B=0.2028{\rm cm}^{-1}$. \cite{grebenev00}
\fi
\clearpage
\newpage
\begin{figure}[ht]
  \includegraphics[height=0.8\linewidth]{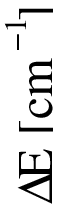}
  \includegraphics[height=0.8\linewidth]{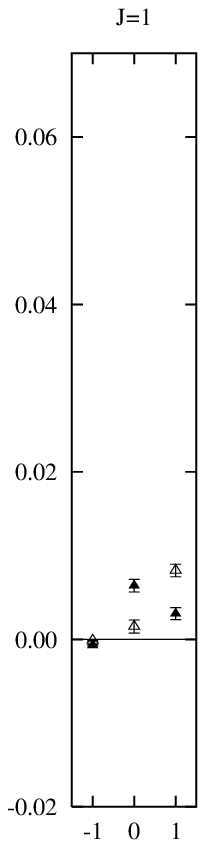}
  \includegraphics[height=0.8\linewidth]{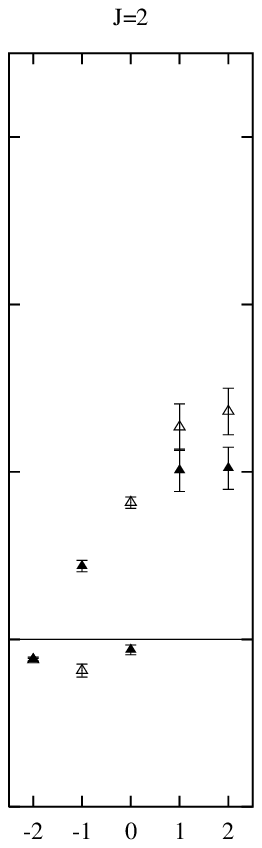}
  \includegraphics[height=0.8\linewidth]{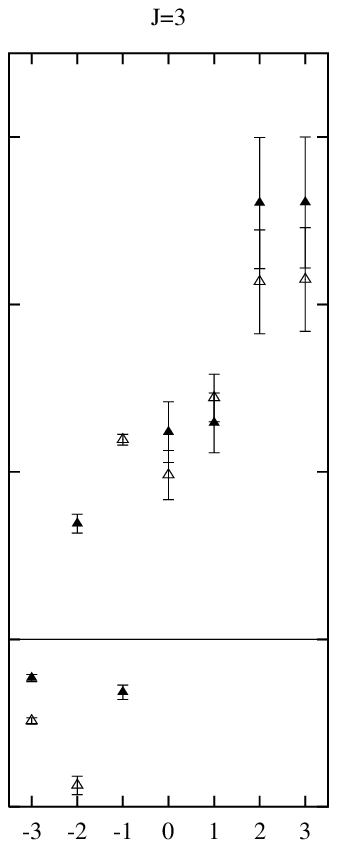}

\vspace*{-0.3truecm}
\hspace*{2.4cm}${\Large\tau}$\hspace*{2.4cm}${\Large\tau}$\hspace*{3.7cm}${\Large\tau}$
\vspace {1truecm}
\end{figure}
\ifcaption
        \noindent Fig.~4a. 
        The error made in ccQA (filled triangles) and EccQA (open triangles),
        $\Delta E_{J,\tau}=
        E_{J,\tau}({\rm (E)ccQA})-E_{J,\tau}({\rm BOUND})$,
        for the complex of OCS with hydrogen, OCS-H$_2$, $J=1,2,3$.
\fi
\clearpage
\newpage
\begin{figure}[ht]
  \includegraphics[height=0.8\linewidth]{DE.eps}
  \includegraphics[height=0.68\linewidth]{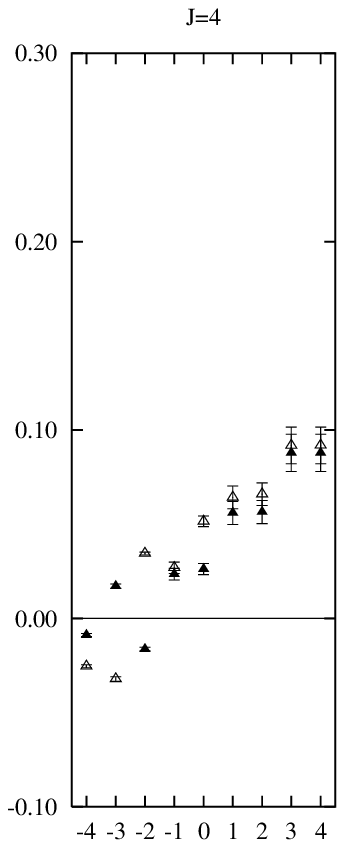}
  \includegraphics[height=0.68\linewidth]{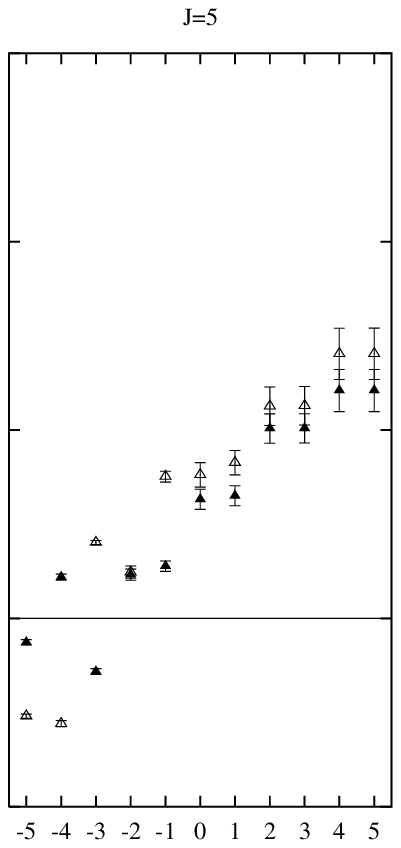}
  \includegraphics[height=0.68\linewidth]{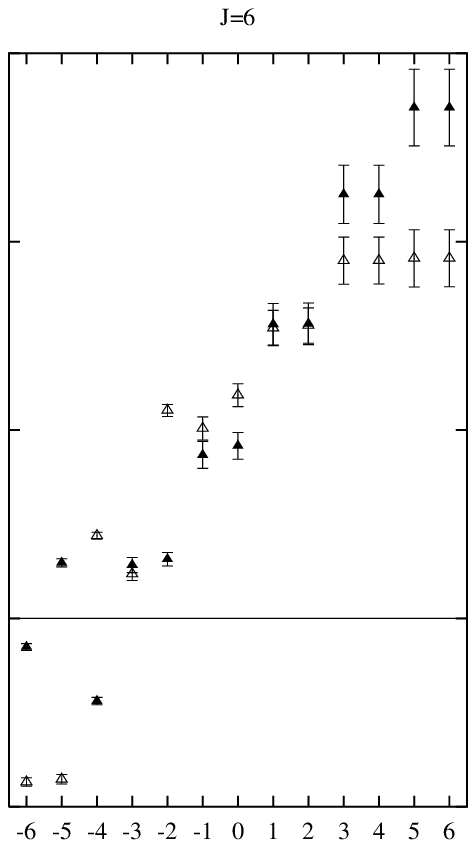}

\vspace*{-0.3truecm}
\hspace*{2.95cm}${\Large\tau}$\hspace*{3.5cm}${\Large\tau}$\hspace*{4.6cm}${\Large\tau}$
\vspace {1truecm}
\end{figure}
\ifcaption
        \noindent Fig.~4b. 
        Same as Fig.~4a for $J=4,5,6$.
\fi
\clearpage
\newpage
\begin{figure}[ht]
  \includegraphics[height=0.8\linewidth]{DE.eps}
  \includegraphics[height=0.8\linewidth]{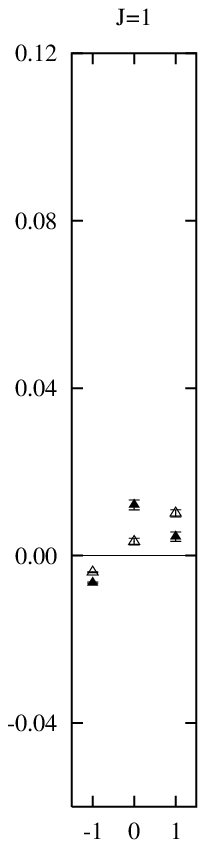}
  \includegraphics[height=0.8\linewidth]{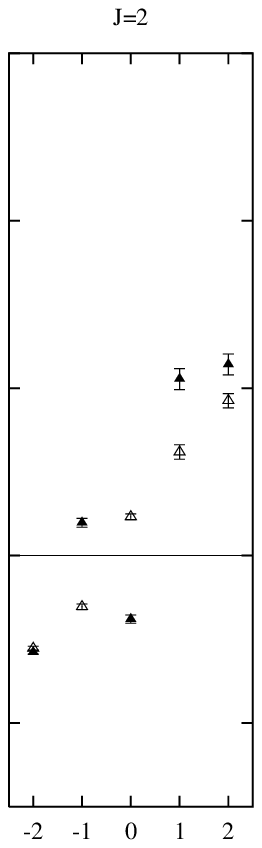}
  \includegraphics[height=0.8\linewidth]{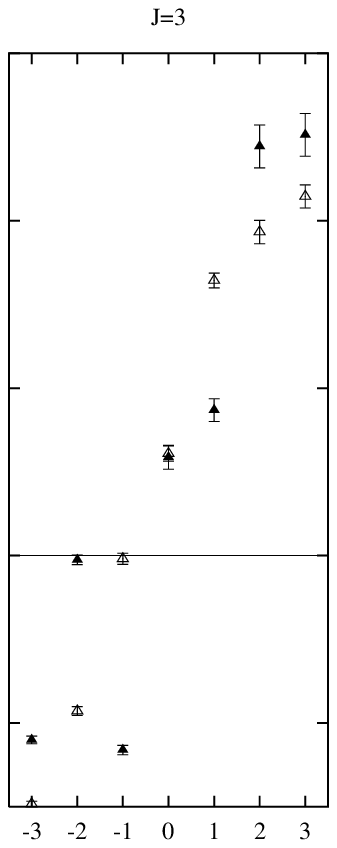}

\vspace*{-0.3truecm}
\hspace*{2.4cm}${\Large\tau}$\hspace*{2.4cm}${\Large\tau}$\hspace*{3.7cm}${\Large\tau}$
\vspace {1truecm}
\end{figure}
\ifcaption
        \noindent Fig.~5. 
        The error made in ccQA (filled triangles) and EccQA (open triangles),
        $\Delta E_{J,\tau}=
        E_{J,\tau}({\rm (E)ccQA})-E_{J,\tau}({\rm BOUND})$,
        for the complex of OCS with helium, OCS-$^4$He, $J=1,2,3$,
        using the MP4 potential of Higgins {\it et.al.\/}~\cite{higgins99}.
	Note the larger vertical scale, relative to that for OCS-(H$_2$) in Fig.~4a.
\fi
\clearpage
\newpage
\begin{figure}[ht]
\includegraphics[height=0.6\linewidth]{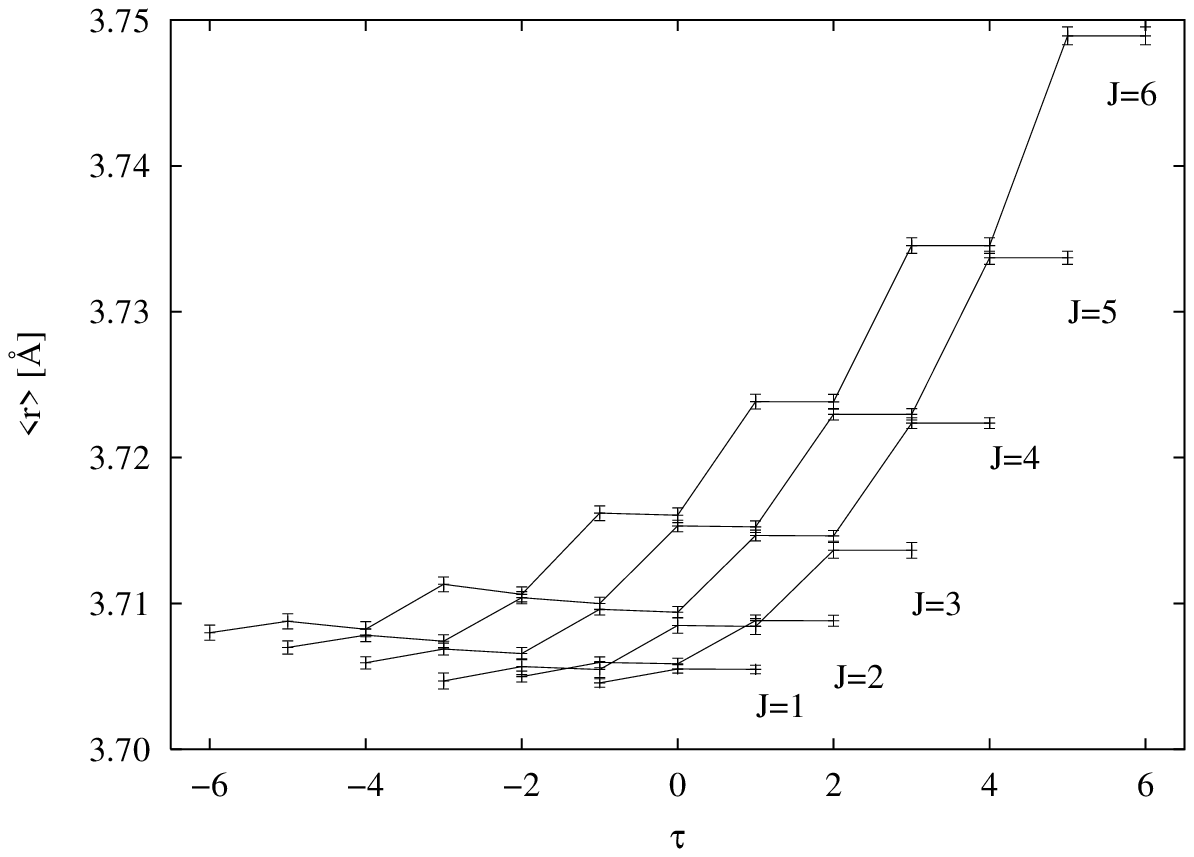}
\includegraphics[height=0.6\linewidth]{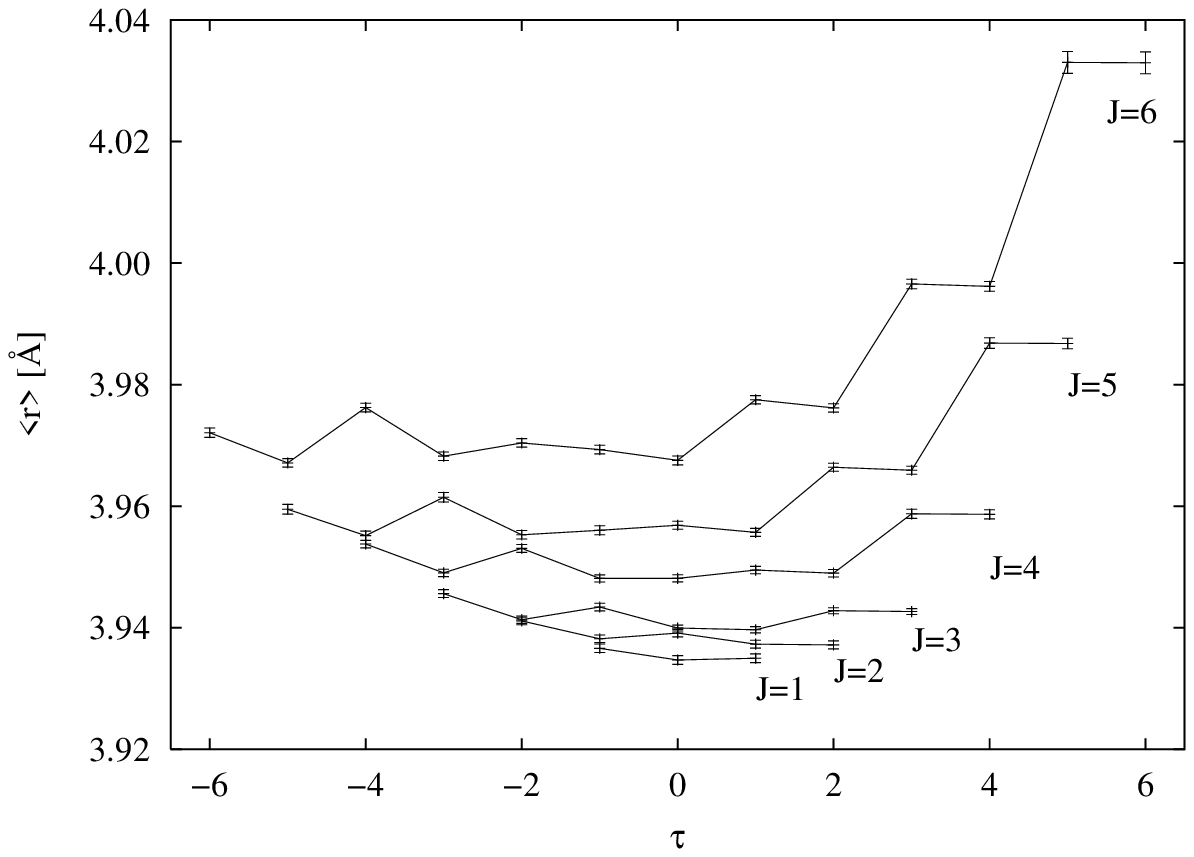}

\vspace {1truecm}
\end{figure}
\ifcaption
        \noindent Fig.~6.  
        Mean distance $\langle r\rangle$ between H$_2$ (top) and
        $^4$He (MP4 potential of Higgins {\it et.al.\/}~\cite{higgins99}, bottom), 
	respectively, and the center of
        mass of OCS, for $J=0,\dots,6$ and $\tau=-J,\dots,J$, in
        ccQA approximation.
\fi
\clearpage
\newpage
\begin{table}
\caption{
        Parameters of the Watson A-reduced Hamiltonian, eq.~\ref{eq:Watson},
        for OCS-(H$_2$) by fitting to the energies obtained by BOUND,
        $J=1,\dots,6$, and to the
        experimental energies of Ref.~\cite{tang02}.
\label{tab:1}
}
\bigskip
\begin{center}
\begin{tabular}{crr}
\hline
& \hspace*{1.5cm} MP4 & \hspace*{1.5cm} exp.\cite{tang02} \\
\hline
$A$ & 0.7595 & 0.7607 \\
$B$ & 0.1997 &  0.19996 \\
$C$ & 0.1540 & 0.15344 \\
$\Delta_J$ & $1.4\times 10^{-6}$    & $2.4 \times 10^{-6}$ \\
$\Delta_{JK}$ & $1.7\times 10^{-4}$ & $1.56 \times 10^{-4}$ \\
$\Delta_K$ & $3.1\times 10^{-4}$    & $10^{-3}$ \\
$\delta_J$ & $3.7\times 10^{-7}$    & $7\times 10^{-7}$ \\
$\delta_K$ & $1.23\times 10^{-4}$   & $1.4\times 10^{-4}$ \\
\hline
\end{tabular}
\end{center}
\end{table}
\clearpage
\newpage
\begin{table}
\caption{
        Parameters of the Watson A-reduced Hamiltonian, eq.~\ref{eq:Watson},
        for OCS-$^4$He by fitting to the energies obtained by BOUND,
        $J=1,\dots,4$, for both, the MP4 and the HHDSD potential, and to the
        experimental energies of Ref.~\cite{tang01}.
\label{tab:2}
}
\bigskip
\begin{center}
\begin{tabular}{crrr}
\hline
& \hspace*{1.5cm} MP4 & \hspace*{1.5cm} HHDSD &
  \hspace*{1.5cm} exp.\cite{tang01} \\
\hline
$A$ & 0.4253 & 0.4309 & 0.44059 \\
$B$ & 0.1833 & 0.1946 & 0.1835992 \\
$C$ & 0.1198 & 0.1191 & 0.1221314 \\
$\Delta_J$    & $6.260\times 10^{-5}$ & $2.244\times 10^{-4}$ &
$3.1378\times 10^{-5}$ \\
$\Delta_{JK}$ & $-5.40\times 10^{-5}$ & $-8.40\times 10^{-5}$ &
$4.669\times 10^{-5}$ \\
$\Delta_K$    & $1.23\times 10^{-3}$  & $2.50\times 10^{-3}$ &
$1.43663\times 10^{-3}$ \\
$\delta_J$    & $1.89\times 10^{-5}$  & $3.42\times 10^{-5}$ &
$1.12265\times 10^{-5}$ \\
$\delta_K$    & $4.38\times 10^{-4}$  & $1.55\times 10^{-4}$ &
$3.2667 \times 10^{-4}$ \\
\hline
\end{tabular}
\end{center}
\end{table}
\clearpage
\newpage
\begin{table}
\caption{
        Parameters of the effective Hamiltonian $H_{\rm rot}^{\rm eff}$,
        eq.~(\ref{eq:Heff1}), for OCS-H$_2$ and OCS-$^4$He
        (MP4 potential of Higgins {\it et.al.\/}~\cite{higgins99}), 
\label{tab:four}
}
\bigskip
\begin{center}
\begin{tabular}{c|r|r}
\hline
& OCS-(H$_2$) [cm$^{-1}$]
& OCS-$^4$He [cm$^{-1}$] \\
\hline
$A$ & 0.7520(2)   & 0.42(3)    \\
$B$ & 0.20318(5)  & 0.206(8)   \\
$C$ & 0.15556(7)  & 0.12084(2) \\
$C'$& 0.15663(5)  & 0.13355(3) \\
$D$ & 0.0016(6)   & -0.07(2)   \\
\hline
\end{tabular}
\end{center}
\end{table}
\clearpage
\newpage
\begin{figure}[ht]
  \includegraphics[height=0.8\linewidth]{DE.eps}
  \includegraphics[height=0.8\linewidth]{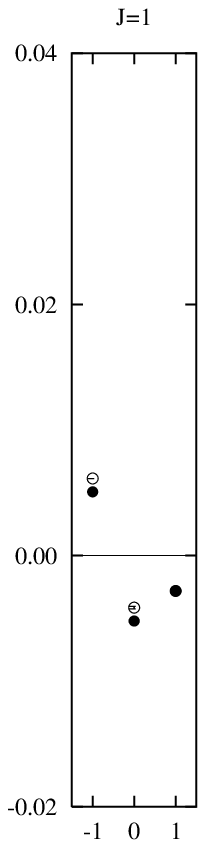}
  \includegraphics[height=0.8\linewidth]{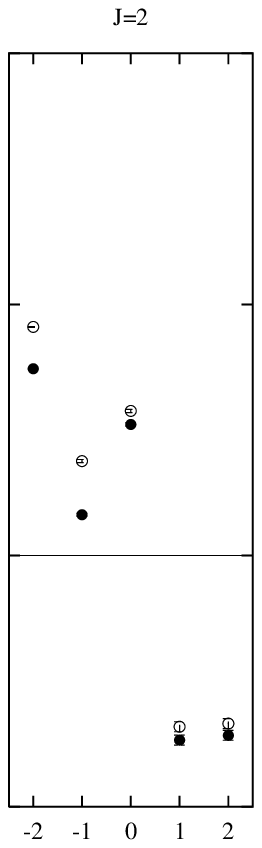}
  \includegraphics[height=0.8\linewidth]{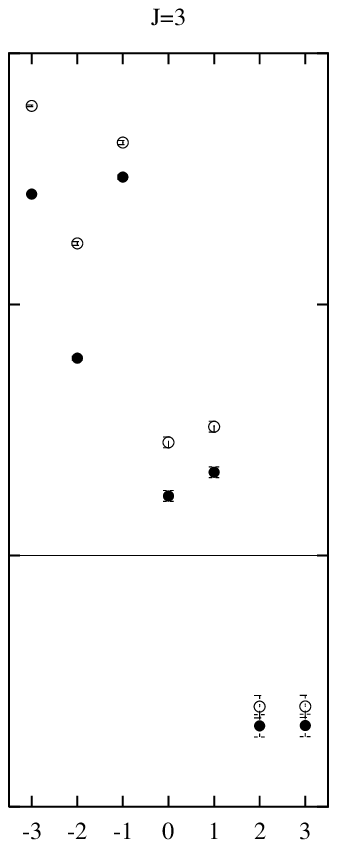}

\vspace*{-0.3truecm}
\hspace*{2.4cm}${\Large\tau}$\hspace*{2.4cm}${\Large\tau}$\hspace*{3.7cm}${\Large\tau}$
\vspace {1truecm}
\end{figure}
\ifcaption
        \noindent Fig.~7a. 
        The error made in the ground state average
        calculation, $\Delta E_{J,\tau}
        =E_{J,\tau}({\rm GSA})-E_{J,\tau}({\rm BOUND})$,
        for the complex of OCS with hydrogen,
	OCS-(H$_2$), $J=1,2,3$. Both inertial matrices, $I$
        (filled circles) and $I'$ (open circles), were used.
\fi
\clearpage
\newpage
\begin{figure}[ht]
  \includegraphics[height=0.8\linewidth]{DE.eps}
  \includegraphics[height=0.68\linewidth]{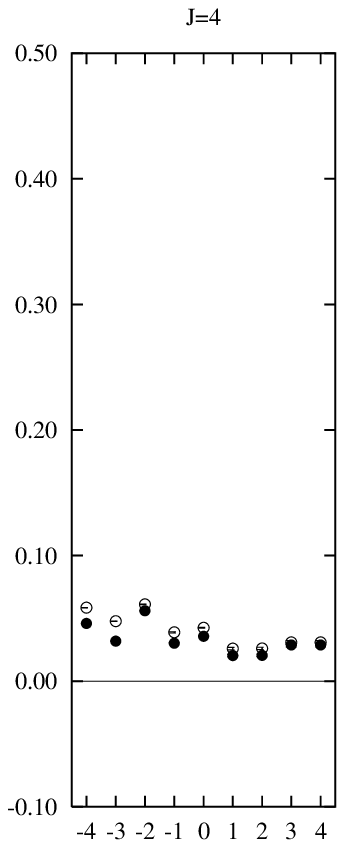}
  \includegraphics[height=0.68\linewidth]{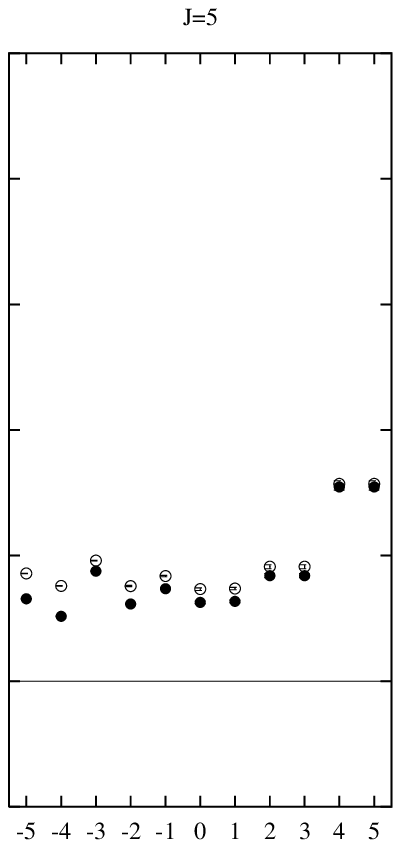}
  \includegraphics[height=0.68\linewidth]{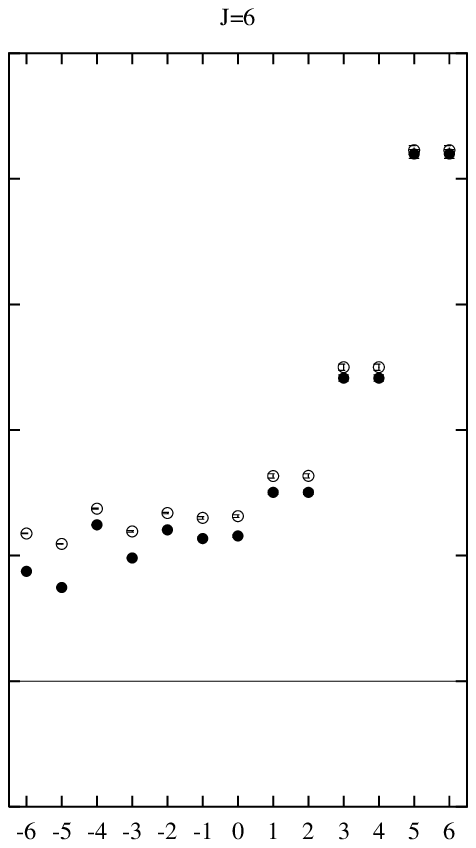}

\vspace*{-0.3truecm}
\hspace*{2.95cm}${\Large\tau}$\hspace*{3.5cm}${\Large\tau}$\hspace*{4.6cm}${\Large\tau}$
\vspace {1truecm}
\end{figure}
\ifcaption
        \noindent Fig.~7b. 
        Same as Fig.~7a for $J=4,5,6$.
\fi
\clearpage
\newpage
\begin{figure}[ht]
  \includegraphics[height=0.8\linewidth]{DE.eps}
  \includegraphics[height=0.8\linewidth]{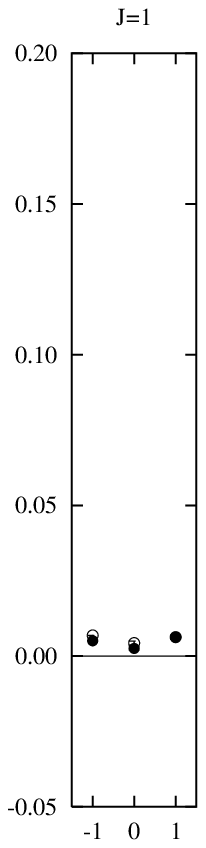}
  \includegraphics[height=0.8\linewidth]{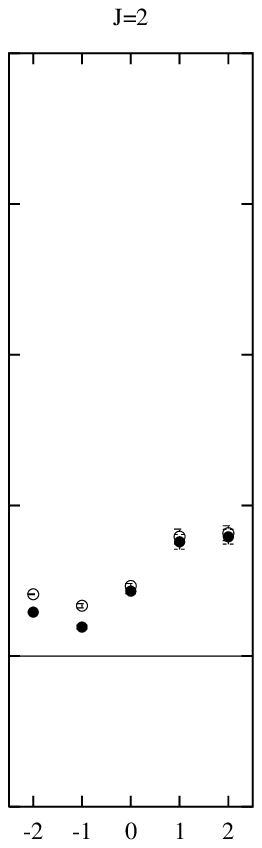}
  \includegraphics[height=0.8\linewidth]{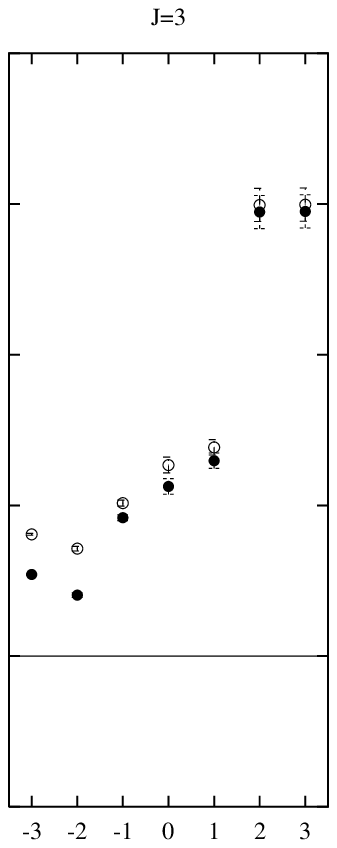}

\vspace*{-0.3truecm}
\hspace*{2.4cm}${\Large\tau}$\hspace*{2.4cm}${\Large\tau}$\hspace*{3.7cm}${\Large\tau}$
\vspace {1truecm}
\end{figure}
\ifcaption
        \noindent Fig.~8. 
        The error made in the ground state average
        calculation, $\Delta E_{J,\tau}
        =E_{J,\tau}({\rm GSA})-E_{J,\tau}({\rm BOUND})$,
        for the complex of OCS with helium,
	OCS-$^4$He, $J=1,2,3$ (MP4 potential of Higgins {\it et.al.\/}~\cite{higgins99}). Both inertial matrices, $I$
        (filled) and $I'$ (open), were used.
\fi
\clearpage
\newpage
\end{document}